\documentclass[%
reprint,
amsmath,amssymb,amsfonts
aps,
superscriptaddress,
prx
]{revtex4-1}

\usepackage{graphicx}
\usepackage{dcolumn}
\usepackage{bm}
\usepackage[colorlinks=true,linkcolor=blue,anchorcolor=blue,citecolor=blue]{hyperref}

\begin{document}

\title{Intrinsic Superflat Bands in General Twisted Bilayer Systems}
\author{Hongfei Wang}
\thanks{These authors contributed equally to this work.} 
\affiliation{Department of Materials Science and Engineering, City University of Hong Kong, Kowloon, Hong Kong 999077, China} 
\author{Shaojie Ma} 
\thanks{These authors contributed equally to this work.} 
\affiliation{Department of Physics, University of Hong Kong, Hong Kong 999077, China}
\author{Shuang Zhang} 
\thanks{shuzhang@hku.hk} 
\affiliation{Department of Physics, University of Hong Kong, Hong Kong 999077, China}
\affiliation{Department of Electrical and Electronic Engineering, University of Hong Kong, Hong Kong 999077, China}
\author{Dangyuan Lei} 
\thanks{dangylei@cityu.edu.hk} 
\affiliation{Department of Materials Science and Engineering, City University of Hong Kong, Kowloon, Hong Kong 999077, China} 
\email[email.address]{nanjing@n}

\date{\today}

\begin{abstract}
Twisted bilayer systems with discrete magic angles, such as twisted bilayer graphene featuring moir\'{e} superlattices, provide a versatile platform for exploring novel physical properties. Here, we discover a class of superflat bands in general twisted bilayer systems beyond the low-energy physics of magic-angle twisted counterparts. By considering continuous lattice dislocation, we obtain intrinsic localized states, which are spectrally isolated at lowest and highest energies and spatially centered around the AA stacked region, governed by the macroscopic effective energy potential well. Such localized states exhibit negligible inter-cell coupling and support the formation of superflat bands in a wide and continuous parameter space, which can be mimicked using a twisted bilayer nanophotonic system. Our finding suggests that general twisted bilayer systems can realize continuously tunable superflat bands and the corresponding localized states for various photonic, phononic and mechanical waves. 
\\
\\
DOI: \href{https://doi.org}{\color{blue}{URL}}
\end{abstract}

\maketitle
\section{Introduction} 
Twisted bilayer systems of two-dimensional (2D) materials, especially for graphene \cite{f1_graphene,f2_graphene} and transition metal dichalcogenides (TMDCs) \cite{f1_TMDs,f2_TMDs}, have recently been employed to explore various physics and applications such as spin-polarized phases \cite{spin1_polarized,spin2_polarized,spin3_polarized,spin4_polarized} and unconventional superconductivity \cite{cond1_graphene,cond2_graphene,cond3_graphene}. 
The structural flexibility further makes twisted van der Waals heterostructures \cite{van1_der,van2_der,van3_der,van4_der} a versatile and tunable platform. 
However, these characteristic behaviors, such as Mott insulating states \cite{mott1_insulator,mott2_insulator,mott3_insulator,mott4_insulator} and superconducting states \cite{spin2_polarized,super1_cond,super2_cond,super3_cond}, are always sensitive to particular discrete twist angles between two sheets, denoted as magic angles \cite{cond2_graphene,magic1,magic2,magic3}, which require high-precision structural manipulation. 
For general twist angles, the scale of moir\'{e} superlattices \cite{super1_lattice,super2_lattice,super3_lattice} ranges in size from unit cells of 2D materials to infinity. 
Although the fragile topology \cite{fragile1,fragile2} for all magic angles has been discussed before, it would be highly desirable to reveal the universal feature of other exotic physical phenomena in general twisted bilayer systems.

At present, moir\'{e} flat bands near the Fermi level underlying the above extraordinary progress have been fully studied \cite{moire_band1,moire_band2}. However, these novel physics and phenomena require precise control of twist angles which are difficult to generalize to distinct artificial materials for various wave systems. General effects of twisted bilayer systems insensitive to twist angles remain out of reach. 

In this work we discover the robust presence of a class of superflat bands in general twisted bilayer systems proved by the tight-binding model (TBM) with negligible next-nearest-neighbor intralayer hoppings. Using the effective macroscopic potential well model (PWM) with spatially modulated couplings, we show that for small twists, localized states definitely appear centered on the AA stacked region (with deepest potential well) at isolated lowest and highest energies, manifesting $C_3$ and $C_6$ symmetries, respectively. Such localized states present neglibile inter-cell coupling, forming superflat bands for general twisted bilayer systems, which is corroborated by exact TBM calculations. We further implement superflat bands and the corresponding localized states via twisted bilayer nanophotonic platforms. Importantly, these superflat bands arise for a continuous set of small angles and do not require fine tuning to the specific magic angles, being readily implementable for various wave systems. 
\begin{figure} 
\includegraphics[scale=0.055]{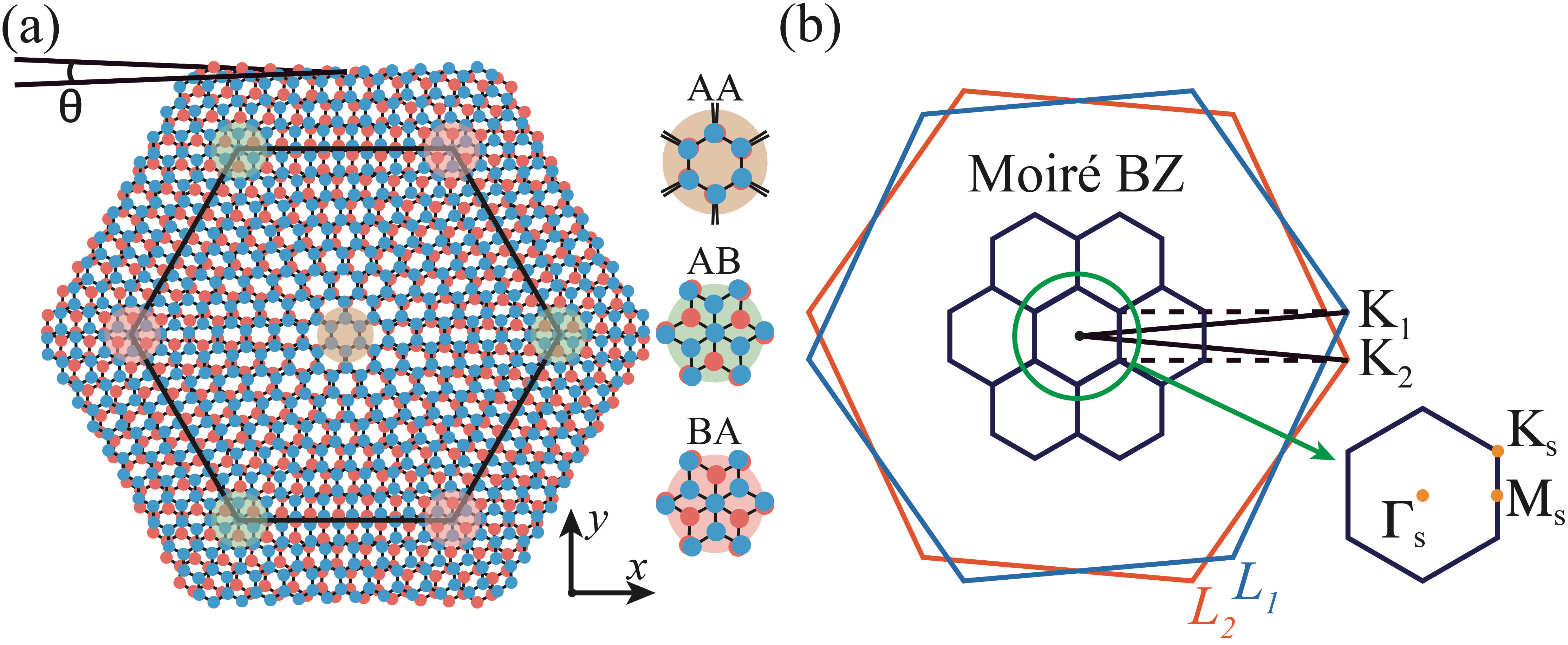}%
\caption{(a) Schematic of general twisted bilayer systems with the twist angle $\theta$, where the largest black hexagon denotes moir\'e superlattices including AA and AB/BA stacked lattices. 
Blue (red) dots represent the sites in $L_1$ ($L_2$). 
(b) Schematic of BZs. Sky blue and red hexagons represent first BZs for $L_1$ and $L_2$ while black hexagons represent moir\'e BZs.} 
\label{fig:1}
\end{figure}

\section{Superflat bands and localized states} 
General twisted bilayer systems display alternating patterns between AA and AB/BA staked lattices (i.e., the A (B) site from the upper layer is perfectly aligned with the A/B (A) site from the lower layer), as illustrated in Fig. \ref{fig:1}(a). In momentum space, rotated unit cells in two layers cause a relative rotation ($\theta$) of first Brillouin zones (BZs), generating an effective moir\'e BZs (see Fig. \ref{fig:1}(b)). 
Periodic moir\'e superlattice has the lattice constant $a_M=\frac{a}{2sin(\theta/2)}$, where $a$ is the lattice constant of primitive unit cells (with a hexagonal p6m symmetry of space groups). 
\begin{figure} 
\includegraphics[scale=0.09]{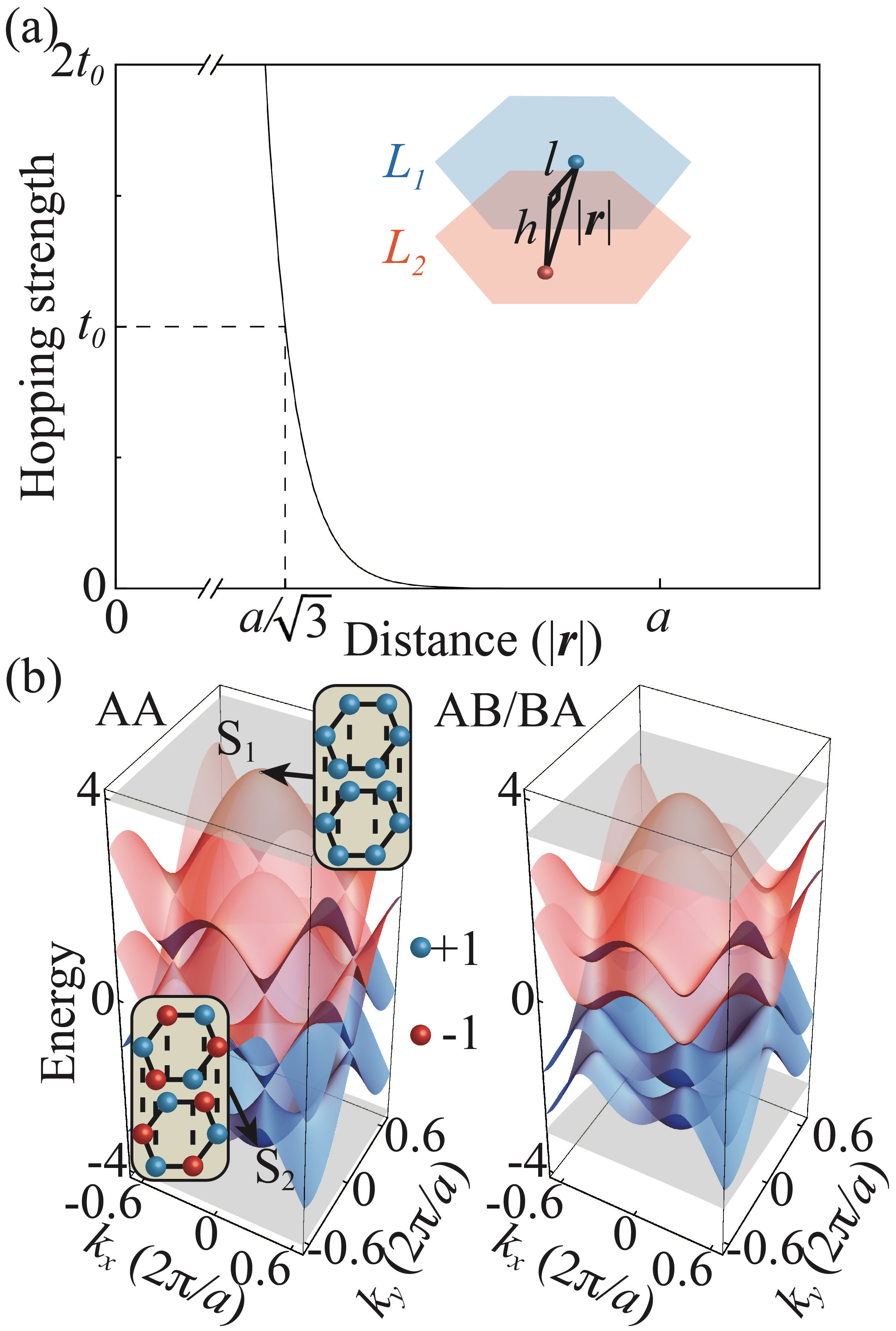}%
\caption{(a) Hopping strength function that decays exponentially with the independent variable of site-to-site distance ${\bm r}$, where only the nearest-neighbor hoppings for intralayer sites are considered. 
(b) Rigorous band structures of AA and AB/BA stacked lattices calculated by the TBM with $h=a/\sqrt{3}$.
The insets show the field distributions of ${\rm S_1}$ (highest energy) and ${\rm S_2}$ (lowest energy) located at ${\rm \Gamma}$ point, implying $C_6$ and $C_3$ symmetries, respectively.} 
\label{fig:2}
\end{figure}
We assume that the hopping rate between every two sites ($i\neq j$) decays exponentially as a function of distance $|{\bm r}_{ij}|$, i.e., $t_{ij}\sim A_0e^{-\gamma |{\bm r}_{ij}|}$, because the classical electronic and photonic systems always allow the overlap of exponential-type wave functions \cite{e1_hopping,e2_hopping,o1_hopping,o2_hopping}. Here $\gamma$ represents the decay rate and $A_0$ is the normalized coefficient constraining the energy scale. In addition, negligible next-nearest-neighbor hoppings of intralayer sites restrict the range of $t_{ij}$ in the following form 
\begin{equation}
\begin{split}
A_{0}e^{-\gamma a/\sqrt{3}}=t_0, A_{0}e^{-\gamma a}\to0. 
\end{split}
\label{eq:1}
\end{equation}
Without losing generality, we set the unit hopping $t_0=1$ in the following analysis. To ensure the dominance of nearest-neighbor hoppings accurately, we further choose $\gamma a\sim30$ corresponding to $t_{ij}(a)\sim10^{-5}\ll t_0$. An exact hopping strength curve is displayed in Fig. \ref{fig:2}(a), where the spatial distance $|{\bm r}_{ij}|=\sqrt{l^2+\rho h^2}$, $l$ and $h$ represent the intralayer and interlayer distances, respectively. $\rho=0$ ($\rho=1$) stands for $i$ and $j$ located at the same (distinct) layers. We model general spinless twisted bilayer systems with the TB Hamiltonian 
\begin{equation}
\begin{split}
H_{\rm TB}=\sum_{\langle i,j \rangle}t_{ij}^{\rho=0}c_{i}^{\dagger}c_j+\sum_{i,j}t_{ij}^{\rho=1}c_{i}^{\dagger}c_j+\sum_{i}\epsilon c_{i}^{\dagger}c_{i}, 
\end{split}
\label{eq:2}
\end{equation}
where $c_i^{(\dagger)}$ corresponds to the creation (annihilation) operator at the site $i$, and $\epsilon$ is the inherent potential which is considered as zero in general systems. This allows us to perform the exact analysis for moir\'e superlattices and provide numerical support for the following detailed models. 
\begin{figure}
\includegraphics[scale=0.078]{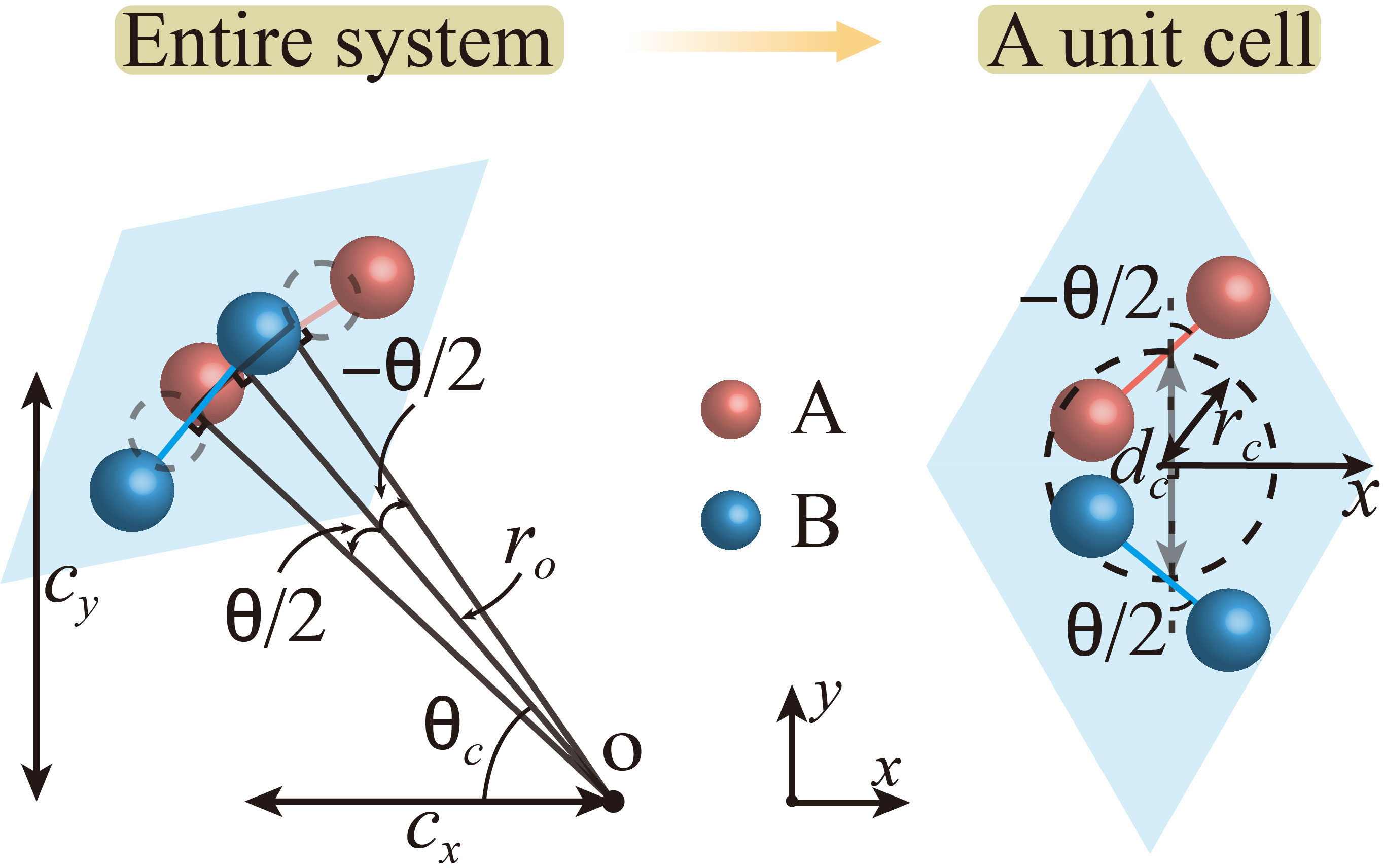}%
\caption{Schematic of the lattice dislocation under specific $r_o$, $\theta$ and $\theta_c$ (left panel), which can be characterized in terms of $\theta_c$, $r_c$ and $d_c$ (right panel).} 
\label{fig:3}
\end{figure}

\begin{figure*}
\includegraphics[scale=0.085]{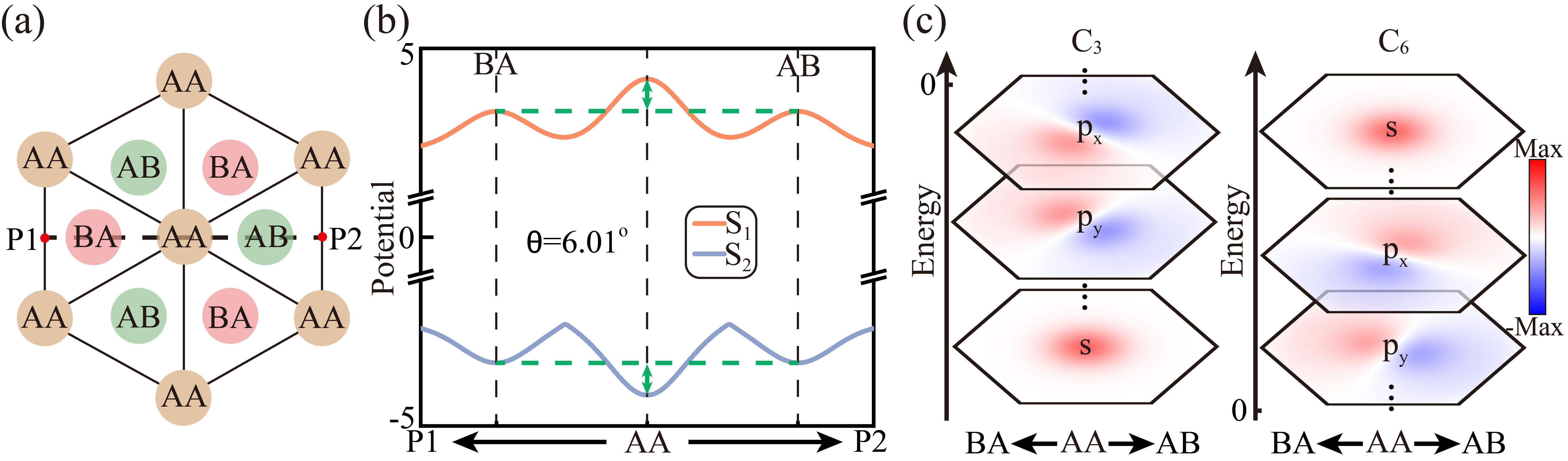}%
\caption{(a) Geometry arrangements of alternating AA and AB/BA stacked lattices and a typical dotted line ${\rm P1P2}$ used for the analysis in (b).
(b) Two potentials of $V$ vary with the spatial parameter between ${\rm P1}$ and ${\rm P2}$ with $\theta=6.01^{\circ}$ and $h=a/\sqrt{3}$, which are obtained by the energies of ${\rm S_1}$ and ${\rm S_2}$. 
Such a description is certainly valid for six different $\theta_c$. 
(c) Calculated eigenstates arranged from the lowest to highest energy. Panels are labeled as $s$ and $p_{x,y}$ states for both $E<0$ and $E>0$ cases. Such filed distributions reveal the corresponding $C_6$ and $C_3$ symmetries consistent with the exact eigenstates at $\Gamma$ point of band structures of AA stacked lattices.} 
\label{fig:4}
\end{figure*}
Furthermore, we calculate the AA and AB/BA stacked band structures under the above hopping relation using the analytical TBM. The lowest/highest energies of first BZs (located at $\Gamma$ point, i.e., the center of blue and red hexagons in Fig. \ref{fig:1}(b)) can be reduced to 
\begin{equation}
\begin{split}
E_{\rm \Gamma}^{\rm AA}&=\pm(t_{ij}(h)+3t_0),\\ 
E_{\rm \Gamma}^{\rm AB/BA}&=\pm\frac{1}{2}(t_{ij}(h)+\sqrt{{t_{ij}(h)}^2+36t_0^2}).\\ 
\end{split}
\label{eq:3}
\end{equation}
Algebraic derivation reveals that for highest (or lowest) bands AA stacked lattices always have higher (or lower) energies than AB/BA stacked lattices, forming a natural potential difference, i.e., $|E_{\rm \Gamma}^{\rm AA}|>|E_{\rm \Gamma}^{\rm AB/BA}|$, unless $h\to+\infty$, that is, $|E_{\rm \Gamma}^{\rm AA}|=|E_{\rm \Gamma}^{\rm AB/BA}|$. Such a relevant energy difference provides a spatial potential well where the deeper potential is located at the AA stacked region with effective masses $m^{*}\sim\pm2\hbar^2/t_0$, with the details given in the Supplemental Material \cite{SM_file}. Here we show a specific case with $h=a/\sqrt{3}$ (see Fig. \ref{fig:2}(b)), where band structures of AA and AB/BA stacked lattices match well with our analysis. Two states (${\rm S_1}$ and ${\rm S_2}$) with highest/lowest energies at $\Gamma$ point of AA stacked lattices present $C_6$ and $C_3$ symmetries, respectively, preserved by irreducible representations in the orthogonal eigenspace, which are the crucial prerequisite for forming superflat bands as following discussions. 

In the vicinity of lowest/highest energies of AA stacked lattices, the previous low-energy theory describing moir\'e bands is invalid \cite{moire_band1}. A concise physical picture can be constructed to depict this system as illustrated in Fig. \ref{fig:3}. The distorted lattices along the azimuth $\theta_c=n\pi/3$, $n=1,2,...,6$, centered around AA stacked lattices, reflect essential characteristics of the potential well. Specifically, for the distorted lattice with a distance from the center of AA stacked region $r_o$, the coordinates of lattice center are $(c_x,c_y)=r_o(cos(\theta_c),sin(\theta_c))$. 
The geometric center of A and B sites is shifted and projected on a specific circle with radius $r_c=2r_osin(\theta/4)$. The distance in x-y plane from one center to another center for two layers is $d_c=2r_osin(\theta/2)$. Here, $r_c$ and $d_c$ are independent of $\theta_c$. In the vicinity of AA stacked region, dislocated lattices for any $r_o$ and $\theta_c$ allow for modelling on a scale of unit cells. A typical case for $n=0$ is displayed in Fig. \ref{fig:3} (right panel). The Hamiltonian around $\Gamma$ point characterizing lattice distortions of the system, $\Phi=\{\phi^1_A,\phi^1_B,\phi^2_{A},\phi^2_{B}\}$, takes the form \cite{gauge1} 
\begin{equation}
\begin{split}
H({\bm k}) = 
\begin{pmatrix}
       h_1&F\\
       F^T&h_2\\
\end{pmatrix},
\end{split}
\label{eq:4}
\end{equation}
where $h_{1,2}=\sigma_x\sum_{i=0}^2t_i-\sigma_y(\pm a\frac{t_1-t_2}{2}k_x+\sqrt{3}a\frac{t_1+t_2}{2}k_y)$ and the wavevector ${\bm k}=\{k_x,k_y\}$. $\sigma_{x,y}$ are the Pauli matrices acting in sublattice space of single layers. $t_1$ and $t_2$ correspond to inter-cell hoppings between A and B sites for single layers along two distinct basis vectors respectively, which are equal for zero $\theta$ or unequal (and exchanged in another layer) for nonzero $\theta$. Besides, the exact derivation manifests that $t_1$ ($t_2$) only grows as $\theta$ decreases (increases), see details in Supplemental Material \cite{SM_file}. The off-diagonal function $F=\{f_{11},f_{12};f_{21},f_{22}\}$ represents the spatially modulated interlayer hoppings, which can be obtained analytically according to Fig. \ref{fig:3} and single depends on $r_o$ under a given $\theta$ \cite{SM_file}. By diagonalizing Eq. (\ref{eq:4}), that is, $E(\Gamma)=PH(\Gamma)P^{-1}$ ($P$ is an invertible matrix), the spatial potential $V(r_o)$ is given by the function $\{min(E(\Gamma),max(E(\Gamma)\}$. In Figs. \ref{fig:4}(a) and \ref{fig:4}(b), we show a specific cross section ${\rm P1P2}$ (with length $\sqrt{3}a_M$) for $\theta_c=0$ or $\pi$, where $\theta=6.01^{\circ}$ and $h=a/\sqrt{3}$. One sees that potential exhibits local valley (peak) characteristic for positive (negative) $m^{*}$ at AA and AB/BA stacked lattices. The potential difference between AA and AB/BA stacked lattices always holds making the central AA stacked lattice become the global extrema of potential. 

\begin{figure*}
\includegraphics[scale=0.072]{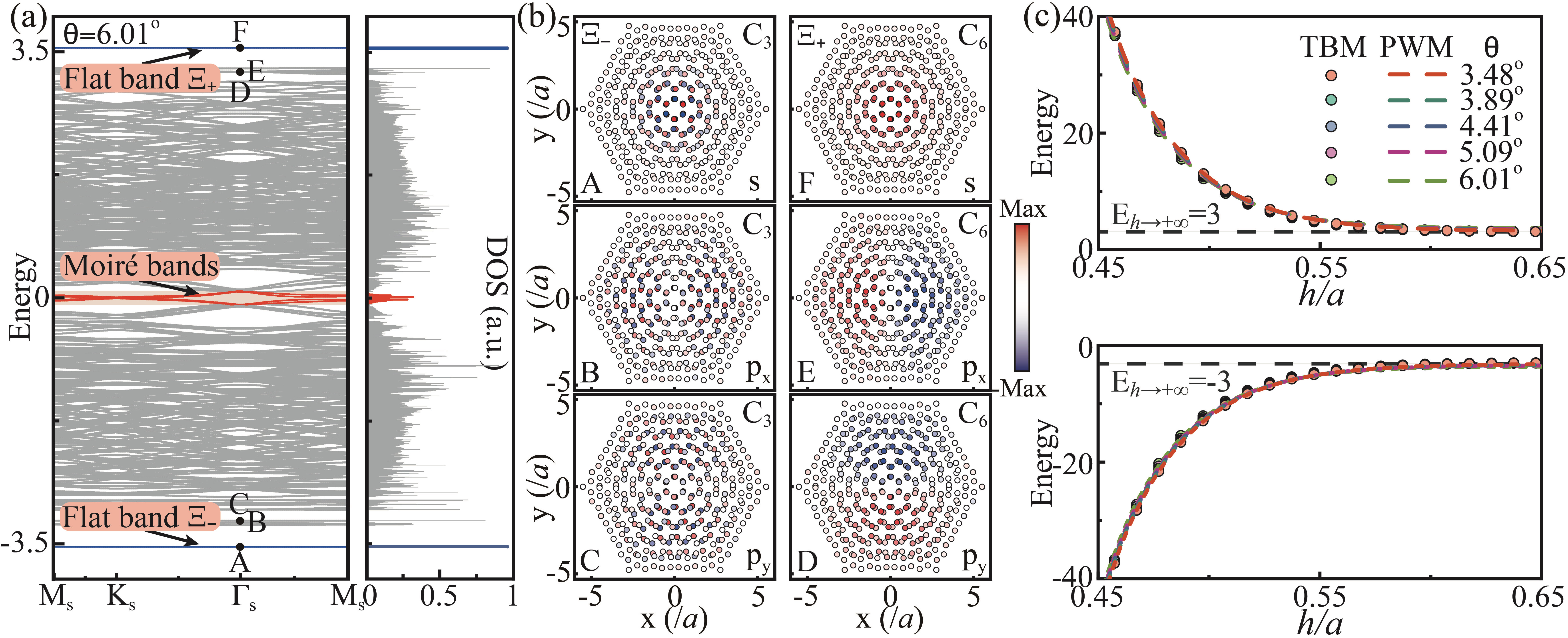}%
\caption{(a) Band structures and related DOS obtained by the TBM with $\theta=6.01^{\circ}$ and $h=a/\sqrt{3}$. Two superflat bands ($\Xi_{-}$ and $\Xi_{+}$) are labeled in blue located at highest and lowest energies with extremely sharp DOS. Besides, typical moir\'e bands labeled in red appear around the zero energy with divergent DOS. 
(b) Various eigenstates for $\Xi_{-}$ and $\Xi_{+}$ bands and adjacent bands, which are arranged from the lowest to highest energy, i.e., A-F, corresponding to $s$, $p_{x,y}$, ..., $p_{x,y}$ and $s$ states. 
(c) The energies of superflat bands varied with $h$ around $a/\sqrt{3}$. Different $\theta$ (i.e., $3.48^{\circ}$, $3.89^{\circ}$, $4.41^{\circ}$, $5.09^{\circ}$ and $6.01^{\circ}$) have also been represented in different colors. Faint circles and dark dotted lines correspond to the results in TBM and PWM, respectively, while faint gray dotted lines indicate the bulk energies of AA stacked lattices for infinite $h$.}
\label{fig:5}
\end{figure*}
Consider the isotropy distortion approximation in the vicinity of central AA stacked region. The system can be regarded as the evolution of a spinless particle with effective mass $m^{*}$ in a given $V(r_o)$ potential well. We describe this process using the time-independent Schr\"odinger-like equation with eigenstates $\Psi$, given by 
\begin{equation}
\begin{split}
[-\hbar^2/2m^{*}(\partial^2_x+\partial^2_y)+V(r_o)]\Psi =E\Psi. 
\label{eq:5}
\end{split}
\end{equation}
The solutions of Eq. (\ref{eq:5}) are shown in Fig. \ref{fig:4}(c). Discrete energy levels correspond to different orders of $\Psi$ manifesting the arrangement of $s$, $p_{x,y}$, ..., $p_{x,y}$, $s$ states from lowest to highest energies. The first half of these states ($E<0$) is composed of ${\rm S}_2$ with $C_3$ symmetry, while the second half ($E>0$) is composed of ${\rm S}_1$ with $C_6$ symmetry. At lowest and highest energies, $s$ states isolated from the continuous bulk energy spectrum exhibit ideal confinement, which can be understood from the confining $V(r_o)$ induced by intrinsic spatial hopping modulations. 

To further demonstrate the properties of general periodic twisted bilayer systems, we calculate band structures of moir\'e superlattices using the TBM with the hopping function of Fig. \ref{fig:2}(a). A representative result for $\theta=6.01^{\circ}$ and $h=a/\sqrt{3}$ is plotted in Fig. \ref{fig:5}(a). Four subbands (red curves) near the zero energy for spinless particles are fully consistent with typical moir\'e bands, corresponding to the divergent density of states (DOS), see the right panel of Fig. \ref{fig:5}(a). Whereas for the lowest and highest energies, superflat bands (blue curves) emerge in isolation accompanied by extremely large DOS, labeled as $\Xi_{-}$ and $\Xi_{+}$. Figure \ref{fig:5}(b) shows typical eigenstates at $\Gamma_{\rm S}$ point of $\Xi_{-}$, $\Xi{+}$ and their adjacent bands. $\Xi_{-}$ (A) and $\Xi_{+}$ (F) correspond to $s$ states formed by ${\rm S}_2$ and ${\rm S}_1$, respectively. The eigenstates for $E<0$ (A-C) and $E>0$ (D-F) cases are consistent with the solution of the above continuous PWM in Fig. \ref{fig:4}(c). We further study the energies of $\Xi_{-}$ and $\Xi_{+}$ with different $h$ and $\theta$ both in TBM and PWM, as displayed in Fig. \ref{fig:5}(c). Since such superflat bands are constrained by the potential of AA stacked lattices, i.e., $E^{\rm AA}_{\Gamma}$, the energies of $\Xi_{-}$ and $\Xi_{+}$ vary exponentially with $h$ in a wide range of $\theta$. As $h\to+\infty$, the energies of $\Xi_{-}$ and $\Xi_{+}$ tend to $-3t_0$ and $3t_0$, respectively, merging into the bulk energy spectrum. 

\section{Nanophotonic implementation} 
\begin{figure*}
\includegraphics[scale=0.095]{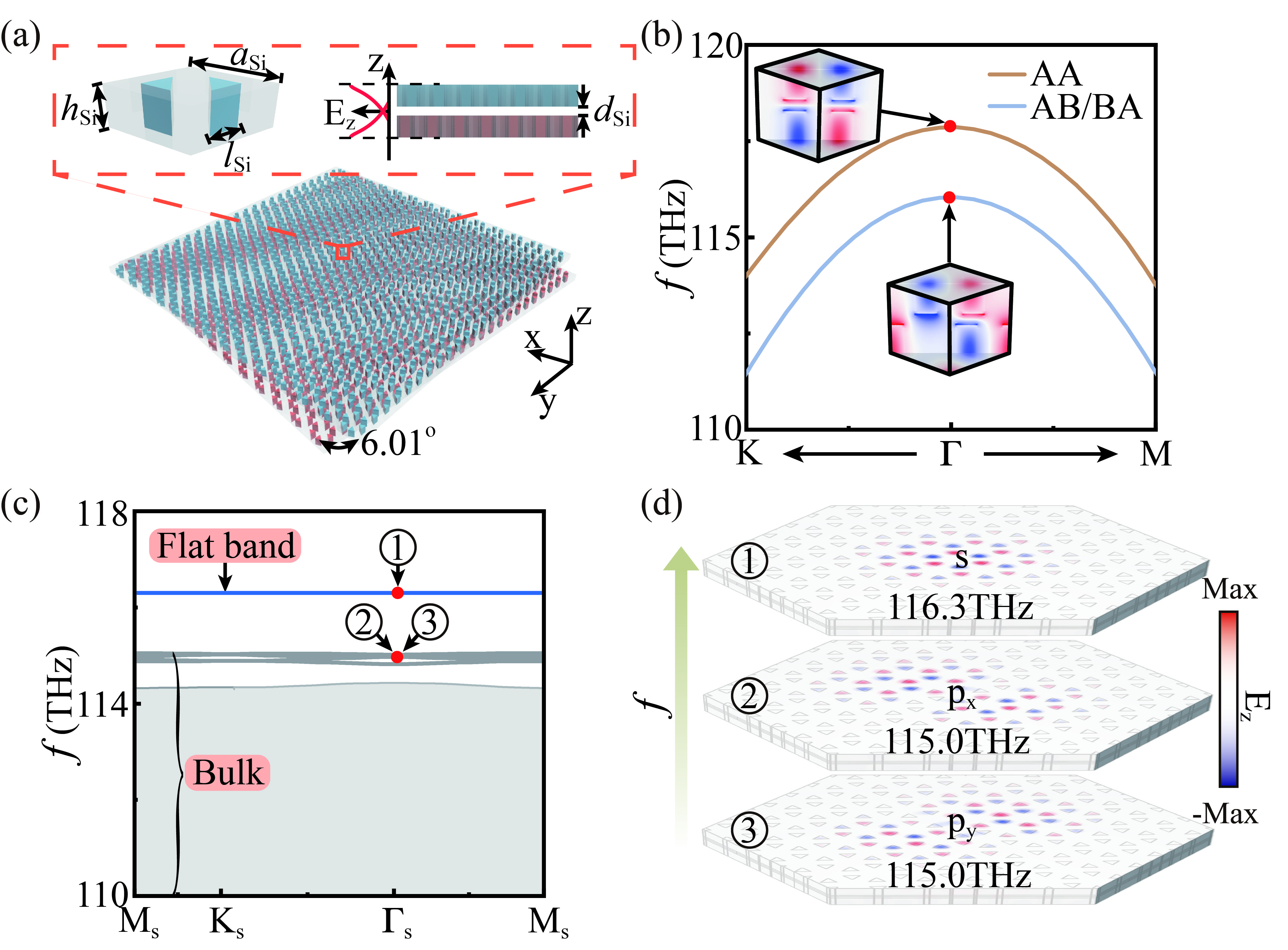}%
\caption{(a) Schematic of twisted bilayer PCs made of silicon and air materials, with graphene-like lattices in each slab. 
The left inset displays the three-dimensional unit cell structure of single layers. The right inset presents the cross section of twisted bilayer PCs and the amplitude (Ez) of fundamental modes along the z direction. 
(b) Band structures of AA and AB/BA stacked PCs near the $\Gamma$ point. The eigenfrequencies for AA stacked PCs are significantly greater than that of AB/BA stacked PCs under the same essential parameters (i.e., $a_{\rm Si}$, $h_{\rm Si}$, $l_{\rm Si}$ and $d_{\rm Si}$). The insets represent the eigenstates with $C_3$ symmetry at the $\Gamma$ point. 
(c) Band structures of moir\'e superlattices with twist angle $6.01^{\circ}$. The superflat band (blue) is separated from the rest of bands. 
(d) Typical eigenstates (Ez) of moir\'e superlattices at ${\rm \Gamma_S}$ point of moir\'e BZs on the superflat band and adjacent bands, i.e., $s$ and $p_{x,y}$ states. $s$ state exhibits well-confined character with $C_3$ symmetries, leading to intrinsic superflat bands. Nonflat bands composed of $p_{x,y}$ states are enumerated as the comparison. Eigenfrequencies and the corresponding electromagnetic fields are solved by COMSOL.}
\label{fig:6}
\end{figure*}

To realize superflat bands and the corresponding localized states in nanophotonic systems, we propose a twisted bilayer photonic crystal (PC) composed of an air layer and two twisted PC slabs, as shown in Fig. \ref{fig:6}(a). Single PC slab has a $C_{6v}$ lattice with lattice constant $a_{\rm Si}=1.5um$ filled with air, where the sublattices are composed of silicon triangular prisms (refractive index $n_{\rm Si}=3.46$) with sidelength $l_{\rm Si}=0.35a_{\rm Si}$ and height $h_{\rm Si}=0.5a_{\rm Si}$ (see the left inset of Fig. \ref{fig:6}(a)). The air layer with a thickness of $d_{\rm Si}=0.2a_{\rm Si}$ is sandwiched between two twisted PC slabs (see the right inset of Fig. \ref{fig:6}(a)). The entire structure is embedded in perfect metal in the stacking direction forming a conservative system (here the transverse magnetic (TM) polarization is considered). 

Owing to the long-wavelength limit of dielectric PCs, we only present the case possessing $C_3$ symmetric states. Figure \ref{fig:6}(b) shows band structures near $\Gamma$ point for AA and AB/BA stacked PCs with given parameters in Fig. \ref{fig:6}(a). The $C_3$ symmetric eigenstates of these two bands preserve particular frequency difference ensuring that the states located in AA stacked lattices is isolated from bulk spectra of twisted bilayer PCs (see the insets of Fig. \ref{fig:6}(b)). Then, we calculate the band structure of twisted bilayer PCs with twist angle $6.01^{\circ}$, as plotted in Fig. \ref{fig:6}(c). The superflat band (blue) is observed at the frequency 116.3THz, describing well-confined $s$ states with $C_3$ symmetry, as shown in the top panel of Fig. \ref{fig:6}(d). Adjacent bands exhibit multipole states of moi\'re superlattices accompanied by worse localization capabilities. For example, $p_{x,y}$ states form crossed and nonflat bands, see Fig. \ref{fig:6}(c) and the middle and bottom panels of Fig. \ref{fig:6}(d). 

Note that such a design process exactly focuses on a single mode of the triangular prism (e.g., the fundamental mode above, which is therefore located in several lower bands). The interaction of different order modes of the triangular prism may merge the superflat bands into upper adjacent bands, which should be avoided when setting essential parameters of the system. 

\section{Discussion} 
The intrinsic superflat bands in our work have the property of isolated energy spectra without mode hybridization between different bands, so that the corresponding eigenstates have a clear and highly symmetrical phase distribution, as shown in Figs. \ref{fig:4}(c) and \ref{fig:5}(c). The localized eigenstates are almost insensitive to periodic moir\'e superlattice boundaries, which is understood as the origin of superflat bands and can be described by the PWM. The carried $C_3$ and $C_6$ symmetries distinguished from moir\'e flat bands formed by the four-band reconstruction (moir\'e bands) near the zero energy have not been fully discussed before \cite{cond1_graphene,cond2_graphene,cond3_graphene,magic1}. Recently, we notice that a displacement electric field is applied in specific twisted bilayer systems (e.g., graphene and boron nitride heterostructure) to study the valley topology of moir\'e bands \cite{valley1,valley2}. In our system, this is equivalent to yield a nonzero $|\epsilon|$ with distinct signs for two layers. The energies of superflat bands will be corrected corresponding to a shift $g(|\epsilon|)$, where $g(|\epsilon|)\ge0$ and grows as $|\epsilon|$ increases, see details given in the Supplemental Material \cite{SM_file}. Apart from that, nonzero $|\epsilon|$ cannot affect the presence of superflat bands and localized states.

\section{Conclusion} 
Combining theoretical PWM analysis and TBM calculation, we have demonstrated a class of superflat bands with $C_6$ and $C_3$ symmetric states for small twists in general twisted bilayer systems. 
The dislocated lattices formed by the systematic hopping modulation create macroscopic effective potential wells centered around the AA stacked region, leading to the well-confined states described by the PWM. 
We also mimic these two effects in nanophotonic systems displaying the unique electromagnetic wave confinement.
Notably, superflat bands and the corresponding localized states can be realized for continuous twist angles (distinct from the discrete set of twist angles in magic-angle physics), showing a class of generalized effects of twisted bilayer systems distinguished from the fragile topology. 
The concept of generalized localized states may inspire a shortcut technology for generating zero-dimensional localization, avoiding complex boundary splicing of (higher-order) topological insulators, which will greatly benefit the wave trapping and manipulation. 
Our results can be extended to photonics \cite{bilayer1_photonic,bilayer2_photonic}, phononics, and mechanical waves, where ideal transport can be realized for integrated chips in information technologies.
 
\begin{acknowledgments}
This work is supported by the Research Grants Council of Hong Kong (CRF Grant No. C6013-18G and PDFS Grant No. PDFS2122-1S04) and the City University of Hong Kong (Project No. 9610434), .
\end{acknowledgments}



\end{document}